\documentclass{article}
\newcounter{rown}
\def\bl{\setcounter{rown}{\value{equation}}
        \stepcounter{rown}\setcounter{equation}0
        \def\theequation{\thesection.\arabic{rown}\alph{equation}}
        }
\def\el{\setcounter{equation}{\value{rown}}
        \def\theequation{\thesection.\arabic{equation}}}

\begin{document}
\renewcommand{\theequation}{\thesection.\arabic{equation}}

\title{On $\kappa$-Deformed $D=4$ Quantum Conformal Group\thanks{To be published in the Proceedings of the
Conference ``Symmetries in Gravity and Field Theory'', held
9-11.06.2003 in Salamanca to celebrate 60-th birthday of Jose A.
de Azcarraga, ed. by V. Aldaya and J.M. Cervero, publ. by
``Ediciones Universidad de Salamanca''.}\thanks{Supported by KBN grant 5PO3B05620}}

\author{Piotr Kosi\'{n}ski$^{(1)}$, Jerzy Lukierski$^{(2)}$ and Pawe{\l}
Ma\'{s}lanka$^{(1)}$
\\ \\
$^{(1)}$Institute of Physics, \\
 University of {\L}\'{o}d\'{z}, ul. Pomorska 149/153,
  \\ 90-236 {\L}\'{o}d\'{z}, Poland
\\ \\
$^{(2)}$Institute for Theoretical Physics,\\ University of Wroc{\l}aw, \\ pl. M. Borna 9,
 50-204 Wroc{\l}aw, Poland}
\date{}

\maketitle

\begin{abstract}
This paper is presented on the occasion of 60-th birthday of
 Jose
Adolfo de Azcarraga
who in his very rich scientific curriculum vitae has also a chapter
 devoted to studies of quantum-deformed symmetries, in particular
  deformations of relativistic
   and Galilean space-time symmetries \cite{sala26}--\cite{sala29}.

  In this paper we provide new steps toward
   describing the $\kappa$-deformed $D=4$ conformal group transformations.
We consider the quantization of $D=4$ conformal group with
dimensionful  deformation parameter $\kappa$. Firstly we discuss
the noncommutativity following  from the Lie-Poisson structure
described by the light-cone $\kappa$-Poincar\'{e} $r$-matrix. We
present complete set of $D=4$ conformal Lie-Poisson brackets and
discuss their quantization. Further we define the light-cone
$\kappa$-Poincar\'{e} quantum  $R$-matrix  in $O(4,2)$
vector representation and discuss the inclusion of noncommutative
conformal translations into the framework of $\kappa$-deformed
conformal quantum group. The problem with real structure
of $\kappa$-deformed conformal group is pointed out.
\end{abstract}
\maketitle

\section{Introduction}

\protect\hskip12pt
The standard $\kappa$-deformation obtained in 1991--92
\cite{sala1}--\cite{sala3} leads to the introduction of quantum
time coordinate. Indeed, after introducing the dual pair of
standard $\kappa$-deformed Poincar\'{e} algebra and
$\kappa$-deformed Poincar\'{e} group \cite{sala4}--\cite{sala7} we
get the $\kappa$-deformed Minkowski space
$\hat{x}_{\mu}=(\hat{x}_i, \hat{x}_0)$ described by the following
algebraic relations.
\begin{equation}\label{sala1.1}
    \left[  \widehat{x}_i , \widehat{x}_j \right] = 0 \, ,
    \qquad
    \left[  \widehat{x}_0 , \widehat{x}_i \right] = \frac{i}{\kappa}
    x_i \, .
\end{equation}
Further in 1995--96 there was introduced the generalized
$\kappa$-deformation \cite{sala8}--\cite{sala10}, with quantized
direction $y_0 = a^\mu x_\mu$ in Minkowski space, where $a_\mu$ is
an arbitrary constant fourvector.  The relations (\ref{sala1.1})
were replaced by the following $a_\mu$-dependent commutator
\cite{sala8}--\cite{sala12}
\begin{equation}\label{sala1.2}
     \left[  \widehat{x}_\mu , \widehat{x}_\nu \right] =
     \frac{1}{\chi}\left(
a^\mu \widehat{x}^\nu - a^\nu \widehat{x}^\mu
     \right)\, .
\end{equation}

In particular if $a_\mu= (1,0,0,0)$ we obtain from (\ref{sala1.2})
the relations (\ref{sala1.1}).

It appears that if $a_\mu$ is light-like ($a_\mu^2 =0$) the
classical $r$-matrix corresponding to the relation (\ref{sala1.2})
satisfies classical Yang--Baxter equation. We call such a
deformation the light-cone quantum $\kappa$-deformation of
Poincar\'{e} symmetries. It has been shown \cite{sala13} that
after nonlinear change of basis the light-cone
$\kappa$-deformation of Poincar\'{e} algebra can be identified
with the null-plane quantum deformation proposed independently by
Ballesteros et all \cite{sala14}.

The classical $r$-matrix (\ref{sala1.2}) if $a_\mu^2 =0$ can be
also used as the classical $r$-matrix describing the quantum
deformation of $D=4$ Weyl and $D=4$ conformal symmetries. It
should be recalled that the light-cone $\kappa$-deformation of
Poincar\'{e} algebra has been extended in \cite{sala8} to the
deformation of $D=4$ Weyl algebra containing besides the
Poincar\'{e} generators $(P_\mu, M_{\mu\nu})$ also the dilatation
generator $D$. Recently \cite{sala11} there was also employed the
light-cone $\kappa$-deformation of $D=4$ Poincar\'{e} algebra in
the framework of twist quantization technique \cite{sala15,sala16}
as a member of three-parameter family of quantum deformations of
$D=4$ conformal algebra.

In this paper we plan to study the $\kappa$-deformation of $D=4$
conformal group, generated by the light-cone $\kappa$-Poincar\'{e}
$r$-matrix. In \cite{sala4} it has been shown that the standard
$\kappa$-deformation of Poincar\'{e} group can be obtained by
quantization of the Lie-Poisson bracket. Such a method has been
extended to the description of noncommutative symmetry parameters
for the $D=4$ Weyl group \cite{sala8} as well as for $D=4$
Poincar\'{e} supergroup \cite{sala17}. Indeed, introducing the
quantum $R$-matrix and applying the FRT method \cite{sala24bis} 
describing noncommutativity by means of ``RTT=TTR equations'' one
can show that quantized Lie-Poisson brackets provide for the
 cases of light-cone $\kappa$-Poincar\'{e} group,
Weyl group and Poincar\'{e} supergroup the transition from
classical to quantum group symmetries. Unfortunately, due to
higher nonlinearities of Lie-Poisson brackets for conformal
translations, such a method does not work in straightforward way
for the $\kappa$-deformed conformal group.

In this paper we consider firstly in Sect. 2 the noncommutativity
of all conformal group parameters obtained from the  Lie-Poisson
structure determined by the light-cone $\kappa$-Poincar\'{e}
$r$-matrix. It appears that in such a way in the presence of
conformal translations
  we obtain only leading terms (of order $\frac{1}{\kappa}$) describing
 the noncommutativity of quantum group parameters. In order to compare
these results with the complete set of quantum group relations, given by
FRT method \cite{sala24bis} providing the noncommutativity in all
orders, we define in Sect. 3 the set of ``RTT =TTR equations'' for
the case of $\kappa$-deformed conformal group. We present
explicite formulae for the conformal quantum symmetry parameters
 which span the coset $\frac{SU(2,2)}{SL(2,C)}$ (translations, conformal
translations and dilatations). In such a framework however
 we were not able to impose the
reality conditions for quantized conformal group parameters. We
would like to recall here that the consistency of quantum
deformations of $D=4$ conformal symmetries with reality structure
($*$-Hopf algebra structure)
is also a nontrivial problem in the discussion of Drinfeld-Jimbo
  $q$-deformation of $SU(2,2) \simeq O(4,2)$ conformal Lie
algebra (see e.g. \cite{sala18}).

Finally it should be mentioned that recently a
$\kappa$-deformation of $D=4$ conformal algebra has been studied
\cite{sala19,sala20} as generated by the Jordanian
$\kappa$-deformation \cite{sala21} of the $O(2,1)$ subalgebra ($D, P_0, K_0$)
 of
$D=4$ conformal algebra. In the paper \cite{sala20}
 there was also pointed out the incompleteness of  quantization
  procedure of conformal
Lie-Poisson bracket relations.
  The complete quantization was only performed in \cite{sala20} for
the Weyl subgroup, and even in this case of simpler deformation
the problem of real ($*$-Hopf algebra) deformation of full $D=4$
conformal group has not been resolved.

\section{Lie-Poisson Structure on $D=4$ Conformal \protect\allowbreak Group
 and its Quantization}
\setcounter{equation}{0}

\protect\hskip12pt
We consider firstly the classical conformal group acting on
generalized space-time with symmetric tensor $g_{\mu\nu}$. The
general conformal transformations on Minkowski space vector
 $u^\mu = (y^i, y^0)$ look as follows:
 \begin{equation}
 \label{sala2.1}
    y^{\mu} = e^{d}\Lambda^\mu _{\ \nu} \frac{y^{\nu}+c^{\nu}y^2   }
    {1+ 2cy +c^2 y^2} + a^{\mu}\, ,
\end{equation}
 where $a^\mu$, d, $c^\mu$ and $\Lambda^{ \nu}_{\ \mu}$
  are
respectively translations, dilatations,
  conformal translations and Lorentz group parameters.

The relevant composition law,
\bl
\begin{eqnarray}
 \label{sala2.2a}
    (a', c', d', \Lambda')\ast (a,c,d,\Lambda)=
    (a'',c'',d'',\Lambda'')\, ,
\end{eqnarray}
reads
\begin{eqnarray}\label{sala2.2b}
\Lambda^{''\mu}_{\quad \nu} &=& \Lambda^{'\mu}_{\ \ \rho}\,
\widetilde{\Lambda}^{\rho}_{\ \sigma} (a,c)\, \Lambda^\sigma_{\
\nu}\, ,
\\ \cr\label{sala2.2c}
a''^{\mu} &= &
 a'^{\mu}
 + e^{d'}
\Lambda^{'\mu}_{\ \ \nu} \, \frac{a^\nu + c'^\nu a^2}{1+2c'a +c'^2
a^2}\, ,
\\ \cr \label{sala2.2d}
c''^\mu &= & c^\mu + e^{d} \Lambda^{\mu}_{\ \, \sigma} \,
\frac{c'^\sigma + a^\sigma c'^2}{1+2c'a +c'^2 a^2} \, ,
\\ \cr \label{sala2.2e}
d'' & = & d + d' - ln \left( 1 + 2c'a + c'^2 a^2 \right)\, ,
\end{eqnarray}
\el
where
\begin{eqnarray}\label{sala2.3}
  \widetilde{\Lambda}^{\mu}_{\ \nu}(a,c) & = &
     \delta^\mu_{\ \nu} + \frac{2}{1+2c'a+c'^2a^2}
     \left(
     -a^\mu c'_\nu +(1+2c'a)c'^\mu a_\nu \right.
     \cr\cr
&&    \left. - c'^2a^\mu a_\nu - a^2 c'^\mu c'_{\ \nu}
     \right)\, .
\end{eqnarray}

It is straightforward to compute the left- and right-invariant
vector fields where $M_{\mu\nu }$ and $P_{\mu}$ are Lorentz and
translations generators respectively. We obtain

i) left-invariant fields:

\begin{eqnarray}\label{sala2.4}
    P_\mu & = & -2c _\mu \frac{\partial}{\partial d } + 2c^\rho
    \left(
\Lambda^\alpha_{\ \rho} \frac{\partial}{\partial \Lambda^{\alpha
\mu}} - \Lambda^\alpha _{\ \mu} \frac{\partial}{\partial
\Lambda^{\alpha\rho}}
    \right)
    \cr
    \cr
    && + c^2 \frac{\partial}{\partial c^\mu} - 2c_\mu c^\alpha
    \frac{\partial}{\partial c^\alpha}
+ e^d \Lambda^\alpha_{\ \mu} \frac{\partial}{\partial a^\alpha}
\cr\cr
M_{\mu\nu} & = & \Lambda^{\alpha}_{\ \mu}
\frac{\partial}{\partial \Lambda^{\alpha\nu}} -
\Lambda^{\alpha}_{\ \nu} \frac{\partial}{\partial
\Lambda^{\alpha\mu}} + \left( c_\mu
\frac{\partial}{\partial c^\nu} - c_\nu
\frac{\partial}{\partial c^\mu}
 \right)
\end{eqnarray}

ii) right-invariant fields:
\begin{eqnarray}\label{sala2.6}
\widetilde{P}_\mu & = & \frac{\partial}{\partial a^\mu}
\cr\cr
\widetilde{M}_{\mu\nu} & = & \Lambda_{\nu\beta}
\frac{\partial}{\partial \Lambda^\mu_{\ \beta}}
-
\Lambda_{\mu\beta}
\frac{\partial}{\partial \Lambda^\nu_{\ \beta}}
+
a_\nu
\frac{\partial}{\partial a^\mu}
- a_\mu \frac{\partial}{\partial a^\nu}\, .
\end{eqnarray}

In order to define the Lie-Poisson structure we select
 the following classical $r$-matrix for the Poincar\'{e} group
  in the generalized space-time with metric $g^{\mu\nu}$
 \begin{equation}\label{sala2.6bis}
    r = \frac{1}{\kappa} M_{0\nu} \wedge P^\nu
\end{equation}
where $P^\nu = g^{\nu\rho} P_\rho$.
The Schouten bracket reads
\begin{equation}\label{sala2.7}
    [r,r] = \frac{g_{00}}{\kappa^2} \, M_{\mu\nu} \wedge P^\mu \wedge P^\nu \, .
\end{equation}

It should be recalled that
 the generalized space-time variables $y_\mu$ can be related with
  physical Minkowski coordinates with metric
   $\eta_{\mu\nu}= diag (1,-1,-1,-1)$
   as follows (see e.g. \cite{sala12})
\bl
   \begin{equation}\label{sala2.8a}
    x_\mu = R_\mu^{\ \nu} \, y_\nu \, ,
\end{equation}
where from the relation $y_\mu g^{\mu\nu} y_\nu = x_\mu
\eta^{\mu\nu} x_\nu$  we get  that
\begin{equation}\label{sala2.8b}
    R^{\mu}_\nu \, \eta^{\nu\tau} \, R_\tau^{\ \rho}
    = g^{\mu\rho} \, , \qquad
    R^\mu_{\ \nu} = \left( R_\mu^{\ \nu} \right)^T  
\end{equation}
\el

 Introducing $a^\mu = R_0^{\ \mu}$  one obtains for  the physical
 Minkowski space symmetries the following formula for the
  classical $r$-matrix (\ref{sala2.6})
  \begin{equation}\label{sala2.9}
    r = a^\mu \, M_{\mu\nu} \wedge P^\nu \, .
\end{equation}
Because from (\ref{sala2.8a}-\ref{sala2.8b}) follows that
 $g^{00} = a^\mu a_\mu$ we obtain that the Schouten bracket
  (\ref{sala2.7}) vanishes if $a^\mu a_\mu= 0$.
In such a case (e.g. if we choose $a_\mu = (1,1,0,0))$ the
classical $r$-matrix (\ref{sala2.9}) satisfies
 the classical YB equation and
  describes the quantum deformation
of any group containing Poincar\'{e} group as its subgroup.
In particular such a classical $r$-matrix has seen
 used in \cite{sala8} to construct the
quantum version of $\kappa$-Weyl group

Below we shall consider quantum deformation of conformal group.
Using Eqs. (\ref{sala2.4}--\ref{sala2.6}) we compute in a standard
way the basic Poisson brackets:
\bl
\begin{eqnarray}\label{sala2.10a}
    \left\{
\Lambda^{\varepsilon}_{\ \lambda} , \Lambda^{\tau}_{\ \rho}
\right\} & = &
- \frac{1}{\kappa}
\left(
2c^\rho \Lambda^\tau_{\ \rho} \left( \Lambda^\varepsilon_{0}
 g_{\lambda\delta}
- \Lambda^\varepsilon_{\ \delta} g_{\lambda 0} \right)
\right.
\cr\cr
&&
- 2c_\delta\left( \Lambda^\varepsilon_{0}\,
\Lambda^\tau_{\ \lambda } - g_{\lambda0} g^{\epsilon\tau}
\right)
\cr\cr
&&
- 2c^\rho\Lambda^\varepsilon_{\ \rho}
 \left( \Lambda^\tau_{0}g _{\lambda\delta}\,
- \Lambda^\tau_{\ \lambda }g_{\delta 0} \right)
\cr\cr
&&
\left.
+ 2c_\lambda \left( \Lambda^\tau_{\ 0} \Lambda^\varepsilon_{\ \delta}
 - g_{\delta 0} g^{\varepsilon\tau}
 \right)
\right)
\end{eqnarray}
\begin{eqnarray}\label{sala2.10b}
\left\{
\Lambda^{\varepsilon}_{\ \lambda} , a^{\tau}
\right\} & = &
- \frac{1}{\kappa}
\left(\Lambda^{\tau}_{\ \lambda}
\left(
e^d \Lambda^{\varepsilon}_{\ 0} - \delta^{\varepsilon}_{0} \right)
+ g^{\varepsilon \tau}
\left(
\Lambda_{0\lambda} - e^d g_{0\lambda} \right)\right)
\end{eqnarray}
\begin{eqnarray}\label{sala2.10c}
\left\{
\Lambda^{\varepsilon}_{\ \lambda} , c^{\tau}
\right\} & = &
- \frac{1}{\kappa}
\left(c^2 \left( \Lambda^{\varepsilon}_{\ 0} \, \delta^\tau_{ \ \lambda}
- \Lambda ^{\varepsilon \tau} g_{\lambda 0} \right)\right.
\cr\cr
&& -2c^\tau c_\lambda
 \Lambda^{\varepsilon}_{\ 0} + 2
\Lambda_{\ \rho}^{\varepsilon}c^{\rho}
\left( c^\rho g_{\lambda 0} - c_0 \delta^{\tau}_{\ \lambda}
  \right)
  \cr \cr
  &&
\left.
  + 2 c_0 \, c_\lambda \Lambda^{\varepsilon \tau}\right)
\cr
 && \\
\left\{
\Lambda^{\varepsilon}_{\ \lambda} , d
\right\} & = &
- \frac{2}{\kappa}
\left(
\Lambda^{\varepsilon}_{\ \rho}\,
c^\rho \, g_{\lambda 0} - \Lambda^{\varepsilon}_{\ 0} c_\lambda \right)
\\ \cr
\left\{
a^{\varepsilon} , a^\tau
\right\} & = &
- \frac{1}{\kappa}
\left( - \delta^\varepsilon_{\ 0}\,  a^\tau + \delta^\tau_{\ 0}\,  a^{\varepsilon}
\right)
\\ \cr
\left\{
a^{\varepsilon} , c^\tau
\right\} & = &
- \frac{1}{\kappa} e^d
\left(
\Lambda^{\varepsilon \tau}\,
c_0 \, - \delta^\tau_0 \Lambda^{\varepsilon}_{\ \rho} c^\rho \right)
\\ \cr
\left\{
a^{\varepsilon} , d
\right\} & = & 0
\\ \cr
\left\{
c^{\varepsilon} , c^\tau
\right\} & = &
- \frac{1}{\kappa} c^2
\left(
c^\tau \, \delta^\varepsilon_0 - c^\varepsilon \, \delta^\tau_{\ 0} \right)
\\ \label{sala2.10i} \cr
\left\{ c^{\varepsilon} , d \right\} & = & - \frac{2}{\kappa}
\left( c^2 \, \delta^\varepsilon_0 - c_0\, c^\varepsilon \right)\,
.
\end{eqnarray}
\el

Eqs. (\ref{sala2.10a}--\ref{sala2.10i}) define Lie-Poisson
structure on conformal group. One can pose the question whether
this structure can be quantized by a ``naive'' correspondence
 principle $\{\cdot , \cdot \}\longrightarrow \frac{1}{i \hbar}[ \cdot , \cdot ]$
  in order
  to obtain a consistent Hopf
algebra structure. By inspecting the
  terms involving conformal translations
  we see that
   after passing by ``naive'' quantization procedure from LP brackets
    to commutators
   we  face the difficulties due to
    the lack of unique
  ordering
 prescription of nonlinear terms.
 This difficulty disappears  if
we consider only the Weyl group. In such a case the ``naive''
quantization of the LP brackets provides in straight-forward way
proper Hopf algebra structure called $\kappa$-Weyl group
\cite{sala8}.

\section{FRT Quantization Technique of the \protect\allowbreak Con\-for\-mal
 Poisson-Lie  \protect\allowbreak Brackets}
\setcounter{equation}{0}

\protect\hskip12pt
 It has been shown in \cite{sala9} that
the $\kappa$-Poincar\'{e} algebra for $g_{00}=0$ case, defined in
\cite{sala8}, is isomorphic to the one introduced by Ballesteros
et all \cite{sala14}. However, for the latter deformation the
quantum $R$-matrix is known \cite{sala22}. This opens the way to
use the FRT approach. Further we shall view therefore
  the conformal group (see
e.g. \cite{sala23}) as the matrix group $SO(4,2)$. It is well
known that the isomorphism between $SO(4,2)$ and the conformal
group can be described as follows.

Let us take the matrix tensor
$g_{AB}$, $A,B=0, \ldots , 5$ in the form
\begin{equation}\label{sala3.1}
g_{AB}=
\left(%
\begin{array}{cccccc}
  0 & 0 & 0 & 1 & 0 & 0 \\
  0 & -1 & 0 & 0 & 0 & 0 \\
  0 & 0 & -1 & 0 & 0 & 0 \\
  0 & 0 & 0 & 0 & 0 & 0 \\
  0 & 0 & 0 & 0 & -1 & 0 \\
  0 & 0 & 0 & 0 & 0 & 1 \\
\end{array}%
\right)
\end{equation}
and consider the cone
\begin{equation}\label{sala3.2}
    g_{AB}\xi^{A} \xi^{B} = 0\, .
\end{equation}
Defining the homogeneous functions on this cone by $(\xi^4 +\xi^5
\neq 0)$
\begin{equation}\label{sala3.3}
    x^{\mu} \equiv \frac{\xi^{\mu}}{\xi ^{4}+\xi^{5}}\, ,
    \qquad \mu = 0,, \ldots, 3
\end{equation}
we find how $SO(4,2)$ acts on $x^\mu$ as conformal group.

Keeping in mind the 
explicit action (\ref{sala2.1})  of conformal group on $x^{\mu}$ and using
(\ref{sala3.1}-\ref{sala3.3}) one can find the parametrization of
general $SO(4,2)$ matrix in terms of conformal transformations.

The general element can be written in the following $6\times 6$
matrix form as the following product ($\mu,\nu = 0,1,2,3;$ $A,B=0,1,2,3,4,5$):
\begin{equation}\label{sala3.4}
    G = T \bullet D\bullet L \bullet C \, .
\end{equation}
The general formulae for the factors in the  decomposition
 (\ref{sala3.4})  looks
as follows:

i) Translations (parameters $a_\mu$)
 \bl
\begin{eqnarray}\label{sala3.5a}
T^A_{\ B} & = & \delta^{A}_{\ B} + a^\mu \delta^{A}_{\ \mu} \left(
\delta^{4}_{\ B} + \delta^{5}_{\ B}\right) + a^\mu g_{\mu B}
\left( \delta^{A}_{\ 4} - \delta^{A}_{\ 5}\right) \cr\cr && +
\frac{1}{2} \, a^2 \left( \delta^{A}_{\ 4} - \delta^{A}_{\
5}\right) \left( \delta^{4}_{\ B} + \delta^{5}_{\ B}\right)
\end{eqnarray}

ii) Dilatations (parameter $\lambda$)

\begin{eqnarray}\label{sala3.5b}
P^A_{\ B} & = & ch\lambda
\left(
\delta^{A}_{\ 5}\delta^{5}_{\ B} +
\delta^{A}_{\ 4}  \delta^{4}_{\ B}
 \right)
 -
 sh\lambda
\left(
\delta^{A}_{\ 5}\delta^{4}_{\ B} +
\delta^{A}_{\ 4}  \delta^{5}_{\ B}
 \right)
\end{eqnarray}

iii) Conformal transformations (parameters $c_\mu$)

\begin{eqnarray}\label{sala3.5abis}
C^A_{\ B} & = & \delta^{A}_{\ B}- c^\mu \delta^{A}_{\ \mu} \left(
\delta^{5}_{\ B} - \delta^{4}_{\ B}\right) + c^\mu
\delta_{\mu\beta} \left( \delta^{A}_{\ 4} + \delta^{A}_{\
4}\right)
\cr\cr
&& - \frac{1}{2} c^2 \left( \delta^{A}_{\ 4} +
\delta^{A}_{\ 5} \right) \left( \delta^{5}_{\ B} - \delta^{4}_{\
B} \right)
\end{eqnarray}

iv) Lorentz transformations (parameters $\Lambda^A_{\ B}$) 

\begin{eqnarray}\label{sala3.5c}
L^A_{\ B} & = & \delta^{A}_{\ 4}\delta^{4}_{\ B}
+
\delta^{A}_{\ 5}\delta^{5}_{\ B}
\cr\cr
&&
+
 \left( 1 - \delta^{A}_{\ 4}\right)
\left( 1 - \delta^{A}_{\ 5}\right)
\left( 1 - \delta^{4}_{\ B}\right)
\left( 1 - \delta^{5}_{\ B}\right)\Lambda^A_{\ B} \, ,
\label{sala3.5cbis}
\end{eqnarray}
\el
 where $\Lambda^\mu_{\ \nu}$ is the Lorentz matrix
corresponding to the $g_{00} = 0$ case.

Using this parametrization  we
write down the classical coproducts which are  equivalent
to the composition law (2.2) with an appropriate ordering: \bl
\begin{eqnarray}\label{sala3.6a}
    \Delta(a^\mu) &= & a^\mu\otimes 1 +
    \left(e^d \Lambda^\mu _{\ \nu} \otimes 1    \right)
     \left(1 \otimes a^\nu + c^\nu \otimes a^2    \right)
\cr
&& \cdot
     \left(1 \otimes 1 + 2 c_\rho \otimes a^\rho + c^2 \otimes a^2 \right)^{-1}
    \cr\cr
     \Delta(e^{-d}) &= & \left( e^{-d} \otimes e^{-d} \right)
     \left( 1\otimes 1 + 2c_\rho \otimes a^\rho + c^2 \otimes a^2\right)
\cr\cr
      \Delta(c^\mu) &= & 1 \otimes c^\mu +
      \left( 1 \otimes 1 +
2 c^\nu \otimes a_\nu + c^2 \otimes a^2
      \right)^{-1}
\cr
&& \cdot
      \left(
c^\beta \otimes 1 + c^2 \otimes a^\beta
      \right)
      \left(
1 \otimes c^d \Lambda^\mu _{\ \beta}
      \right)
      \cr\cr
      \Delta(\Lambda^\mu_{\ \nu}) &= & \Lambda^\mu_{\ \alpha}
      \otimes \Lambda^\alpha_{\ \nu}
      + 2 \left(
\Lambda^\mu_{\ \alpha} c^\alpha \otimes a_\beta \Lambda^\beta_{\ \nu}
      \right)
      \cr\cr
&&      - 2 \left( \Lambda^{\mu}_{\ \alpha} \otimes 1 \right)
\left( c^\alpha \otimes a^2 + 1 \otimes a^\alpha \right)
\left( 1 \otimes 1 + 2c_\rho \otimes a^\rho + c^2 \otimes a^2 \right)^{-1}
\cr\cr
&& \cdot
\left( c_\beta \otimes 1 + c^2 \otimes a_\beta \right)
\left( 1 \otimes \Lambda^\beta_{\ \nu}\right)
\end{eqnarray}
\el

Now we introduce noncommuting conformal group parameters.
According to Ref. \cite{sala22} the quantum $R$-matrix is obtained
by the following exponentiation procedure
\begin{eqnarray}\label{sala3.7}
R & = &  e^{ \frac{i}{\kappa} M^{32} \otimes P_{2} }
\, e^{ \frac{i}{\kappa} M^{31} \otimes P_{1} }
\, e^{ - \frac{i}{\kappa} P_{0} \otimes M^{30} }
\, e^{ \frac{i}{\kappa} M^{30} \otimes P_{0} }
\cr\cr
&& \cdot \ e^{ - \frac{i}{\kappa} P_{1} \otimes M^{31} }
\, e^{ - \frac{i}{\kappa} P_{2} \otimes M^{32} }\, .
\end{eqnarray}
The computation of matrix realization of (\ref{sala3.7}) is
simplified due to the nilpotency of translation generators. We
obtain
\begin{equation}\label{sala3.8}
    R = I \otimes I + \frac{i}{\kappa}\, R_1 + \frac{i}{\kappa^2}\,
    R_2 \, ,
\end{equation}
where
\bl
\begin{eqnarray}\label{sala3.9a}
(R_1)^A_{\ B}{}^{C}_{\ D} & = &
\left[
\left( \delta^A_4 - \delta^A_5\right)\delta^2_B
-
\delta^2_A \left( \delta^4_B + \delta^5_B\right)
\right]
\left( \delta^C_0 \delta^2_D + \delta^C_2 \delta^3_D \right)
\cr\cr
&& +
\left[
\left( \delta^A_4 - \delta^A_5\right)\delta^1_B
-
\delta^A_1 \left( \delta^4_B + \delta^5_B\right)
\right]
\left( \delta^C_0 \delta^1_D + \delta^C_1 \delta^3_D \right)
\cr\cr
&&+
\left(  \delta^A_0 \delta^0_B - \delta^A_3\delta^3_B \right)
\left[\delta^C_0
\left( \delta^4_D + \delta^5_D\right)
+ \delta^3_D
\left( \delta^C_4   - \delta^C_5  \right)\right]
\cr\cr
&& + \left[
 \delta^A_0\left( \delta^4_B + \delta^5_B\right)
+
\delta^3_B\left( \delta^A_4 -  \delta^A_5\right)
\right]
\left( \delta^C_3 \delta^3_D - \delta^C_0 \delta^0_D \right)
\cr\cr
&&  +
\left( \delta^A_0 \delta^1_B + \delta^A_1 \delta^3_B \right)
\left[\delta^C_1
\left( \delta^4_D + \delta^5_D\right)
 -
 \delta^1_D \left( \delta_4^C - \delta^C_5\right)
 \right]
 \cr\cr
 && +
\left( \delta^A_0 \delta^2_B + \delta^A_2 \delta^3_B \right)
\left[
\delta^C_2
\left( \delta^4_D + \delta^5_D\right)
- \delta^2_D
 \left( \delta^C_4 - \delta^C_5\right)
 \right]
\end{eqnarray}

\begin{eqnarray}\label{sala3.9b}
(R_2)^A_{\ B}{}^{C}_{\ D} & = &
\left( \delta^A_4 - \delta^A_5\right)
\left( \delta^4_B + \delta^5_B\right)
+
\delta^A_0 \, \delta^3_B
\left( \delta^C_4 - \delta^C_5\right)
\left( \delta^4_D + \delta^5_D\right)
\cr\cr
&& -
\left( \delta^A_4 - \delta^A_5\right)
\delta^3_B \, \delta^C_0
\left( \delta^4_D + \delta^5_D\right)
\cr\cr
&&
- 2
\delta^A_0
\left( \delta^4_B + \delta^5_B\right)
\left( \delta^C_4 - \delta^C_5\right)
\delta^3_D \, .
\end{eqnarray}
\el
Using the basic relation of FRT formalism \cite{sala19}
\begin{equation}\label{sala3.10}
    R^A_{\ B}{}^C_{\ D} \, G^B_{\ E} \,  G^D_{\ F}
    =
    G^A_{\ B}\, G^C_{\ D}
    R^B_{\ E}{}^D_{\ F} \, ,
\end{equation}
we compute the commutation rules which determine the quantum conformal
group algebra. This calculation is quite involved because, due to the complicated
 parametrization of $SO(4,2)$ matrices, and we present only the relations
  for the parameters from the coset $\frac{SU(2,2)}{Sl(2;C)}$.

  We get
  \bl
\begin{eqnarray}\label{sala3.11a}
\left[
\widehat{a}{\ }^\mu , \widehat{a}{\ }^\nu \right] & = & \frac{i}{\kappa^2} \left(
\delta^\nu_0 \, \widehat{a}{\ }^\mu -
\delta^\mu_0 \, \widehat{a}{\ }^\nu
\right)
\\ \label{sala3.11b} \cr
\left[ e^{-\widehat{d}} , \widehat{a}{\ }^\mu \right] & = & 0
\\ \cr
\left[
e^{-\widehat{d}} , \widehat{c}{\ }^\mu \right] & = &
\frac{2i}{\kappa}
\left(
\delta_0^{\ \mu}
\widehat{c}_\alpha e^{-\widehat{d}} \widehat{c}{\ }^\alpha
- \widehat{c}_0 e^{-\widehat{d}} \widehat{c}{\ }^\mu
\right)
\cr
&&
-\frac{2}{\kappa^2} \, e^{-\widehat{d}} \widehat{c}{\ }^2 \,
\widehat{c}_0 \delta_0^{\ \mu}
\\ \label{sala3.11c} \cr
\left[ e^{-\widehat{d}} \widehat{c}{\ }^\mu , e^{-\widehat{d}}
\widehat{c}{\ }^\nu \right] & = & \frac{i}{\kappa}
e^{-\widehat{d}}
\widehat{c}{\ }^2 e^{-\widehat{d}} \left( \delta_0^{\ \alpha}
\widehat{c}{\ }^\mu - \delta_0^\mu
\widehat{c}{\ }^\nu \right)\, , \label{sala3.11d}
\end{eqnarray}
and more complicated relation
\begin{eqnarray}\label{sala3.11e}
\left[
\widehat{a}^\mu , \widehat{c}_\nu \right] & = & \frac{i}{\kappa}
\left( \left(
\widehat{c}{\ }^\alpha
\widehat{\Lambda}{\ }^\mu_{\ \alpha} + 2
\widehat{c}{\ }^\alpha
\widehat{a}{\ }^\mu e^{-\widehat{d}}
\widehat{c}_\alpha
\right)g_{0\nu}
-
\widehat{c}_0  \widehat{\Lambda}{\ }^\mu_{\ \nu} - 2
\widehat{c}_0
a{\ }^\mu e^{-\widehat{d}}
\widehat{c}_\nu \right)e^{\widehat{d}}
\cr\cr
&&
+ \frac{2}{\kappa^2}
\left(
\widehat{c}_\nu \widehat{\Lambda}{\ }^\mu_{\ \beta}
\widehat{c}{\ }^\beta
+
\widehat{c}_0
\widehat{a}{\ }^\mu e^{-\widehat{d}}
\widehat{c}{\ }^2
- e^{-\widehat{d}}
\widehat{c}{\ }^2\, \widehat{\Lambda}{\ }^\mu_{\ 0}
\right.
\cr
&& \left.
-
2 e^{-\widehat{d}}
\widehat{c}{\ }^2
\widehat{a}{\ }^\mu e^{-\widehat{d}}
\widehat{c}_0 \right)
e^{\widehat{d}} g_{0\nu}
\cr\cr
&&-
\widehat{a}{\ }^\mu
\left(
\frac{2i}{\kappa}
\left(
g_{0\nu}
\widehat{c}_\alpha e^{-\widehat{d}}
\widehat{c}{\ }^\alpha -
\widehat{c}_0 e^{-\widehat{d}}
\widehat{c}_\nu
\right)
\right.
\cr
&&
\left.
- \frac{2}{\kappa^2}
e^{-\widehat{d}}
\widehat{c}{\ }^2
\widehat{c}_0 g_{\nu 0}
\right)e^{\widehat{d}}\, .
\end{eqnarray}
\el

We see that the noncommutative translations $a_\mu$ form the
algebra of $\kappa$-Minkowski space. Further the  elements
$e^{-d}$ and $c_\mu$ span another closed subalgebra. It can be
rewritten in slightly more transparent form if we define
\begin{equation}\label{sala3.12}
    \widehat{s}\equiv e^{-\widehat{d}}
    \widehat{c}{\ }^2 \, , \qquad
    \widehat{l}{\ }^\mu \equiv e^{-\widehat{d}}
    \widehat{c}{\ }^\mu \, .
\end{equation}
Then we obtain
\begin{eqnarray}\label{sala3.13}
    \left[ \widehat{s}, e^{-\widehat{d}} \right] & = & 0 = \left[
    \widehat{s},
    \widehat{l}{\ }^\mu \right] \, ,
    \cr\cr
 \left[ e^{-\widehat{d}},
 \widehat{l}{\ }^\mu \right] & = &
 \frac{2i}{\kappa}
 \left(
\delta_0^{ \ \mu}
\widehat{l}{\ }^2 -
\widehat{l}_0
\widehat{l}{\ }^\mu
 \right)
 - \frac{2}{\kappa^2} \, \widehat{s} \,
 \widehat{l}_0 \, \delta_0 ^{\ \mu}\, ,
 \cr\cr
 \left[
 \widehat{l}{\ }^\mu ,
 \widehat{l}{\ }^\nu \right] & = & \frac{i}{\kappa} \, \widehat{s}
 \left(
 \delta_0 ^{\ \nu}\,
 \widehat{l}{\ }^\mu -
  \delta_0 ^{\ \mu}\,
  \widehat{l}{\ }^\nu
 \right)\, .
\end{eqnarray}
The relations (\ref{sala3.13}) we shall call $\kappa$-deformed
conformal translations algebra.

In the relations (\ref{sala3.11a}-\ref{sala3.11e}) we describe the
noncommutativity of quantum group parameters from the coset
$\frac{O(4,2)}{O(3,1)}\simeq \frac{SU(2,2)}{Sl(2;C)}$. By some
additional calculational effort one could complete this list by
including the noncommutativity relations involving quantum the parameters
$\widehat{\Lambda}{\ }^A_{\ B}$.

Unfortunately, already set of relations
(\ref{sala3.11a}-\ref{sala3.11e}) imply difficulties in
introducing the structure of real quantum conformal group. It can
be checked, that the hermicity properties of quantum conformal
translations imply that $\widehat{c}_\mu \widehat{c}{\ }^\nu=0$;
   this
 constraint is not invariant under the action of coproduct
(\ref{sala3.6a}).
We conclude that our quantum conformal group can not be
 formulated as real quantum algebra.

\section{Conclusions}

\protect\hskip12pt
It should be pointed out that defining noncommutative version of
conformal symmetry is physically quite strongly motivated. Classical
conformal symmetry implies that there is no geometrical scale
(elementary length) which enters  the conformal geometry describing
  conformal space-time
structure. On other hand the notion of classical space-time seems
to be restricted only to the distances larger than the Planck length
(see e.g. \cite{sala24}). Recently it is often conjectured (see e.g.
\cite{sala30}--\cite{sala32})
 that  the Planck length
parameter should enter  the noncommutative  geometry
 taking  into consideration the quantum gravity effects.
  The quantum
conformal symmetries with deformation parameter $\kappa$ could be
 an example of
 such noncommutative  symmetries which describe
   algebraically  the quantum effects at the sub-Planck
distances (in particular $\kappa$ can be identified with the
Planck mass).

\end{document}